## Economic Struggles and Inflation: How Does that affect voting decision?

Muhammad **Hassan** Bin Afzal
Department of Political Science and Public Service
University of Tennessee at Chattanooga
WPF218@Tennessee.edu
https://orcid.org/0000-0001-8192-0885

**Disclaimer:** This is a work in progress, and I would greatly appreciate any feedback, suggestions, or collaborating invitations. Thank you for your time and for reading my effort.

## Abstract:

Economic hardships significantly affect public perception and voting intentions in general elections. The primary focus of my study is to capture the degree of influence that individual economic hardships have on their voting. I utilize the ANES 2024 Pilot Study[1] Survey dataset and introduce a novel composite Inflation Behavior Index (IBR) that captures individuals' cumulative economic and cost of living experience. To that effect, the primary objectives of the current study are threefold: first, to develop a composite economic behavior index from available data and variables to capture the overall economic experience of U.S. individuals due to ongoing inflation; second, to examine how this economic behavior impacts political engagement and voting behavior utilizing appropriate and fitting mathematical models; and finally which specific personal experiences and perceptions about economy and cost of living likely to revoke party loyalty in upcoming U.S. presidential election. My study finds that increased personal economic struggles (pocketbook voting) due to inflation make it more likely for individuals to express an intention to vote against the Incumbent even if the Incumbent is from their self-identified political party.

Conversely, having a negative perception of the national economy (sociotropic voting) is less likely to revoke party loyalty in the upcoming General election. In simpler terms, voters are more likely to vote along party lines even if they perceive their party (the Incumbent) is not handling the economy and cost of living well. Having a higher level of Education, they are more likely to vote for the incumbents in both pocketbook and sociotropic scenarios. Therefore, the current research shows that party loyalty during general elections often persists but is expected to be undermined if the voter experiences adverse economic conditions due to inflation in the past. The findings of the current study provide tools and resources to craft agendas, policies, and strategies not only for policymakers and campaign strategists but also for the political parties to take focused and evidence-based actions to protect and ensure the economic well-being of the general population as well as increase the likelihood to perform better in elections and public opinion measures in a highly polarized political environment.

---

[1] American National Election Studies. 2024. ANES 2024 Pilot Study [dataset and documentation]. March 19, 2024 version. www.electionstudies.org.

**<u>Introduction</u>**

      The effect of inflation on the general population in the United States has a profound and lasting impact on civic engagement and political behavior. Regardless of their ideological alignment, major news outlets such as Fox News[2], CNN[3], and ABC News[4] Report inflation severely impacted the general population in 2023 (Kornick 2024; Wallace 2024; Zahn 2024). An extensive collection of socio-economic and public opinion research shows that economic conditions influence voter behavior.[5] in democracies (Michael Steven Lewis-Beck and Nadeau 2011; Michael S Lewis-Beck and Paldam 2000; Armutcu and Tan 2021), with voters less likely to vote for incumbents during economic downturns and more likely to vote for them during better financial situations. In the context of United States elections, research shows that voters tend to be sociotropic and retrospective[6] In their voting behavior during presidential elections (Lanoue 1994; Markus 1992). However, voters often make prospective choices when the Incumbent is not running.[7], focusing on future economic expectations (Nadeau and Lewis-Beck, 2001). This phenomenon was famously highlighted by James Carville's slogan for Bill Clinton's 1992 campaign: "It's the economy, stupid!" (Benedictis-Kessner and Warshaw 2020). Traditionally, economic circumstances have strongly influenced electoral outcomes, with voters often less likely to vote for incumbents during economic downturns and more likely to reward them during prosperous times (Lewis-Beck and Nadeau, 2011).

      Economic voting theory suggests that voters' choices are influenced by the economic conditions they experience. To operationalize the cumulative financial struggles of the general

---

[2] According to Fox News, 72% of Americans cited inflation has negatively affected their grocery purchases (Kornick 2024).

[3] According to a recent CNN report, almost 65% of US consumers said price inflation negatively impacted their financial well-being in 2023 (Wallace 2024).

[4] According to ABC News, analysts have seen slowing economic growth and persistent inflation despite a healthy economy, creating concerns about future financial stability (Zahn 2024).

[5] Economic voting, a pivotal concept in political science, highlights the substantial influence of current economic conditions on voter choices during elections. Research consistently demonstrates that voters are more likely to support the incumbent when the economy is robust, a trend observed in democracies worldwide. This underscores the crucial role of our research in understanding and predicting election outcomes.

[6] Sociotropic means thinking about how the whole economy is doing , retrospective means considering how the economy has performed in the past, and prospective means assessing how the economy will perform in the election cycle.

[7] Prospective voting involves voters making decisions based on expected future economic conditions rather than past performance.

population, I developed a composite index that captures four different behaviors that someone experienced in the past due to financial constraints. My goal through this research is to examine the cumulative experience of inflation in the daily lives of individuals and how that shapes civic engagement in terms of voting in the general election and how the intersectional sociodemographic characteristics and political party subscription affect their intention to vote and who to vote in the upcoming 2024 U.S. Presidential election. Notably, during crises, the legislative policy entrepreneurs[8] utilize the crisis as a policy window to advance policies less relevant to the ongoing situation and more likely to be associated with personal and party ambitions. (Afzal 2021; 2022b; 2022a). Therefore, my research study examines how personal economic experience influences the intention to vote in general elections. I want to emphasize that the primary focus of my research is to build prediction models to forecast the outcome of elections. Instead, I tried to capture inflation's profound and lasting implications on daily lives and how these economic strains' influence may influence the intention to vote and who to vote in presidential elections. The ANES 2024 Pilot Survey[9] is very fitting in this context because the survey asks about four behaviors that could have occurred in the past due to high inflation, such as borrowing money to pay bills, cutting down on spending, postponing big purchases, and finally dipping into savings. In the current study, I propose the Inflation Behavior Index (IBR) to examine the degree of economic constraints and how that impacts economic decisions, intention to vote, and who to vote in the upcoming U.S. presidential election in November 2024.

**Theoretical Framework: Economic Voting**

The underlying theoretical framework for my study posits that economic conditions strongly influence voting choices in U.S. presidential elections. More specifically, personal economic struggles strongly shape voting behavior and override party loyalty if the cumulative economic struggles are too high over the past year. On the other hand, a personal understanding that the incumbent party is not handling the U.S. economy and cost of living well does not defect party loyalty, and voters are more likely to vote along their party lines even if they think their party is not handling the national economy and cost of living well.

---

[8] Policy entrepreneurs seek to shape the policy process and its results through strategic or gradual actions (Kingdon 1984; Mintrom 1997; Shearer 2015).
[9] The ANES 2024 Pilot Study, funded by the National Science Foundation and conducted by YouGov, is a cross-sectional survey designed to test new questions for potential inclusion in the ANES 2024 Time Series Study (ANES 2024).

The PollyVote project has shown the value of incorporating diverse econometric predictive models applied to elections in the U.S. and Germany, demonstrating that combining forecasts can improve accuracy over individual models. (Graefe 2022). Moreover, the political-economy model by Jérôme, Jérôme-Speziari, and Lewis-Beck (2021) has effectively predicted German election outcomes by considering both economic and political factors (Jérôme, Jérôme-Speziari, and Lewis-Beck 2022). Cross-national studies also support the idea that economic conditions affect voting behavior (Weschle 2014). For example, research in Germany and Canada shows that voters hold governments accountable for financial performance. (Alford and Legge 1984; Happy 1986). Similarly, in the U.K., economic perceptions significantly influence voting behavior. (Tilley, Garry, and Bold 2008).

Several econometric models and socio-behavioral analyses find that individuals tend to hold the government responsible for their economic distress and increasing cost of living, and their dissatisfaction tends to reflect in their voting behavior in general elections. Inflation and higher cost of living significantly negatively impact the Incumbent's chance to be reelected in general elections (Kayser and Peress 2012; Hibbs, Rivers, and Vasilatos 1982; Anderson 2007). Various economic performance indicators such as GDP, inflation rate, and employment status are also employed to assess the perception of voters and how likely that impacts their voting choice in general elections (Michael Steven Lewis-Beck and Nadeau 2011; Michael S Lewis-Beck and Paldam 2000; Tilley, Garry, and Bold 2008). Table 1 summarizes research studies that capture the effect of economic voting in the context of U.S. Presidential elections. This table demonstrates the varying impacts of economic conditions on voting behavior, both when the Incumbent is running and when they are not.

**Table 1:  Summary of Economic Voting Effects**

| | Incumbent Running | | Incumbent Not Running | |
|---|---|---|---|---|
| | Econ: Good | Econ: Not Good | Econ: Good | Econ Not Good |
| Pocketbook Voting | Voters are likely to support the Incumbent if their economic situation improves. (Wu and Huber 2021; Kim and Yang 2022; Sinha et al. 2022) | Voters are less likely to support the Incumbent if their economic situation worsens. (Wu and Huber 2021; Kim and Yang 2022; Sinha et al. 2022) | Voters focus on future economic expectations and prefer candidates promising continued economic growth. (Sinha et al. 2022; Ahmed and Pesaran 2022; Lockerbie 2023) | Negative future economic expectations lead to support for candidates promising change or improvement. (Sinha et al. 2022; Ahmed and Pesaran 2022; Lockerbie 2023) |
| Sociotropic Voting | Incumbents benefit from favorable national economic conditions like low unemployment and economic growth. (Ahmed and Pesaran 2022; Guntermann, Lenz, and Myers 2021; Liao 2023) | Poor national economic conditions result in decreased support for the Incumbent. (Liao 2023; Guntermann, Lenz, and Myers 2021; Ahmed and Pesaran 2022) | Favorable national economic conditions favor candidates from the Incumbent's party. (Ahmed and Pesaran 2022; Guntermann, Lenz, and Myers 2021; Liao 2023) | Poor national economic conditions lead to a preference for opposition candidates promising economic recovery. (Liao 2023; Ahmed and Pesaran 2022; Guntermann, Lenz, and Myers 2021) |
| Retrospective Voting | Voters reward incumbents for positive past economic performance. (Song 2022; Sorace 2021) | Incumbents are punished for negative past economic performance. (Song 2022; Sorace 2021) | N/A | N/A |
| Prospective Voting | N/A | N/A | Voters focus on future economic expectations and prefer candidates promising continued economic growth. (Sinha et al. 2022; Ahmed and Pesaran 2022; Lockerbie 2023) | Negative future economic expectations lead to support for candidates promising change or improvement. (Sinha et al. 2022; Ahmed and Pesaran 2022; Lockerbie 2023) |

Table 1 outlines and distinguishes various forms of economic voting in elections and explores voting intentions and behavior overall. When the Incumbent is running for re-election, and the prospective voter's experience at a personal level (pocketbook voting) and overall perception of the national economy is good, they are more likely to vote for the Incumbent for re-election. Conversely, suppose the economy is not good at a personal and perceived national level. In that case, the prospective voters are less likely to vote for the incumbent re-election.

The table's N/A (not applicable) entries showcase certain cases where specific economic voting types do not apply. For example, when the Incumbent is not running for re-election, judging the past economic situation to form a voting intention and opinion (retrospective economic voting) would be impractical. For the current research, I aim to capture the effects of pocketbook and sociotropic voting, considering when the incumbent president aims to rerun for election. I hypothesize.

**H₁:** An increase in personal economic struggles (pocketbook) decreases the likelihood of voting for incumbent re-election.

**H₁ₐ:** An Increase in personal economic struggles (pocketbook) increases the likelihood of voting against the party line.

**H₂:** Perception of the national economy and cost of living not doing well (sociotropic) decreases the likelihood of voting for the incumbent re-election.

**H₂ₐ:** Perception of the national economy and cost of living is not doing well (sociotropic) increases the likelihood of voting against the party line.

The following section discusses the data cleaning and data collection process in detail. After the reference section, the Codebook provides a detailed guide on how each variable is coded, cleaned, and operationalized in the current study.

**Data and Methods:**

I run an appropriate and fitting set of six regression models to capture the effects of pocketbook and sociotropic economic voting in forming the intention to vote among individuals who experience the impact of inflation and financial struggles. I used the ANES 2024 Pilot Study, which the American National Election Studies conducted. The ANES survey involved approximately 1,779 participants and was conducted online from February 20th to March 1st, 2024. Weights were applied to ensure that the sample accurately represents the broader U.S. population, accounting for demographic factors such as age, gender, race, and Education (ANES 2024). I also share the Codebook after the reference section at the end of this document to summarize all the variables used in this study, the cleaning and coding process and mechanism, and the multicollinearity check. I use the variable that asks the participants whom they would likely vote for in the 2024 election as the primary dependent variable (DV) in all six models.

The rationale behind using this variable as the main DV across all six models is to capture the effect of pocketbook and sociotropic economic voting intention in the upcoming general US presidential election. As of July 26th, 2024, the incumbent U.S. President decided not to run for re-election and endorsed Vice President Kamala Harris as the potential presidential candidate from the Democratic Party facing ex-President Trump for the upcoming November 2024 election. (Bose and Hunnicutt 2024; Spady 2024). The main objective of my research is not to explicitly capture whether current President Biden would have won the 2024 re-election. My primary goal is to examine the profound effects of economic crisis and inflation on U.S. voters and how they express their voting intention and perception of the national economy. The DV also aligns well with the pocketbook and sociotropic economic voting theoretical framework when the Incumbent decides to run for the presidential election. Finally, when the survey took place, it was almost certain that incumbent President Biden from the Democratic party and previous President Trump from the Republican party would run for the 2024 general election. Therefore, the DV captures the voting intention for the Incumbent in the 2024 U.S. Presidential election, where a vote for Biden is coded as 1 and a vote for the non-incumbent, Trump, is coded as 0. The primary independent variable for Models 1,2 and 3 focuses on pocketbook economic voting.

I operationalized the cumulative personal financial struggles in the past through the novel Inflation Behavior Index to capture the tendency of pocketbook voting behavior. The composite index measures cumulative personal economic struggles over the past year. I constructed it from four binary variables (1 = Yes; 0 = No): whether individuals borrowed money to pay bills, cut down on everyday spending, canceled or postponed major purchases, and dipped into savings. The index ranges from 0 to 4, with 0 indicating no inflation-related behavior and 4 indicating engagement in all four behaviors. I also checked the internal consistency of the index, which was validated using Cronbach's Alpha (0.5672), and its construct validity was confirmed through Exploratory Factor Analysis (EFA) and Confirmatory Factor Analysis (CFA), revealing strong factor loadings and a single-factor structure explaining 43.64% of the variance. The Inflation Behavior Index, validated through EFA and CFA, shows strong factor loadings, affirming its construct validity despite a moderate Cronbach's Alpha of 0.5672.

The primary independent variable for Models 4, 5, and 6 focuses on capturing the effect of sociotropic economic voting, which means how the voters perceive how the political parties would handle the economy better. I provided the detailed coding scheme and histogram at the end of the reference. The original variable asks, "Please tell us which political party— the [Democrats/Republicans] or the [Republicans/Democrats]—would do a better job handling each of the following issues, or is there no difference (The cost of living and rising prices)." There are a total of 1298 responses, and 527 explicitly select the republican party as the political party that would do a better job in handling the cost of living and rising prices. I relabel this variable: Do you think the democratic party is not handling the economy and raising prices better? Where yes is 527 and no is the rest of the responses. However, I also ran a missing case loop before running all the regression models to keep the total responses to each variable uniform (the total number of responses to each variable is 1176).

I also control gender, logged age, Education, family income, employment status, residential area, home ownership, and party identification in models 2 and 3 (pocketbook economic voting) and, similarly, models 5 and 6 (sociotropic voting). I stay consistent with well-known studies on economic voting (Hall and Yoder 2019; C. Lewis-Beck and Martini 2020; Becher and Donnelly 2013; Guntermann, Lenz, and Myers 2021). Most existing economic voting studies mainly focus on GDP and macroeconomic conditions. However, my current study primarily focuses on recent self-expressed economic experiences due to rising inflation and cost of living. That is why I also control several sociodemographic variables that affect an individual's voting behavior (intention to vote whom, to be exact).

I include age, income, level of Education, working status, homeownership, and rural/urban residence. Age is a significant predictor due to the widely known influence of age on political preferences. I also include income and education levels to examine how these factors shape an individual's political views and voting behavior. Several studies also emphasize the importance of employment status, which indicates one's economic status and, in most cases, is very likely to determine a person's political opinion. Homeownership status also provides a look at one's economic status, which may have more significant consequences for voting. The dichotomy of living in an urban area versus a rural area is often a true reflection of the political views of these people. However, I chose not to incorporate race as a predictor in my models. Although race may profoundly impact voting behavior, several prominent voting behavior studies did not include race in their economic voting research.

Additionally, the dependent variable of my study is primarily focused on socio-economic factors. It would have brought multicollinearity with some of the included variables. Therefore, adding race as a factor could open up multifaceted interactions and subtleties that are not possible for my study to address satisfactorily and effectively. However, I would like to add race to my future economic models to make them more inclusive and insightful for future research potential.

Finally, the pocketbook full model (Model 3) also includes an interaction term between a binary self-identified Democrat (1 =Yes; 0= No) and interacts with the novel composite Inflation Behavior Index (IBR). My primary motivation for this interaction variable is to examine, even if someone identifies with the incumbent party, their likelihood of voting intention for the Incumbent in the upcoming general election if they experience various degrees of inflation behavior. This interaction term (dem4IBD) specifically addresses my hypothesis ($H_{1a}$), which examines, in the case of pocketbook economic voting, whether varying degrees of inflation-associated behavior likely defect party loyalty in the general elections.  Similarly, I also introduce another interaction term between being a Democrat and thinking about whether the Democratic Party is handling the economy and the cost of living better or not (dem_not_econ_dem). I use the second interaction variable in my model 6 for a full sociotropic voting scenario to capture my hypothesis ($H_{2a}$). The primary objective of these two interaction variables is to capture the degree of varying inflation behavior and sociotropic case, the party loyalty likely to defect, and the potential voter expressing a clear intention to vote against their party candidate.

**Findings and Discussions**

I use six regression models to investigate economic voting conditions under which party loyalty defects in U.S. Presidential elections. The first three models focus on pocketbook economic voting, while the last three examine sociotropic economic voting. Each set of three models starts with simple binomial logistic regression with odds ratio, then adds the control variables, and finally adds the interaction variable to capture specific fracture in party loyalty defect in the upcoming general election.

Model 1, a bivariate logistic regression, assesses the direct relationship between the Inflation Behavior Index and voting for Biden, highlighting the effect of personal economic struggles. Model 2 adds control variables like gender, logged age, Education, family income, employment status, residential area, home ownership, and party identification, isolating the effect of personal economic hardships. Model 3 introduces interaction terms to explore how party loyalty might crack under personal financial difficulties, particularly among Democrats experiencing inflation-related behaviors.

The same structure is used to explore sociotropic economic voting. Model 4 examines the relationship between the perception that Democrats are not handling the cost of living well and voting for Biden. Model 5 includes the same control, providing a clearer picture of national economic dissatisfaction. Finally, model 6 introduces interaction terms to analyze how party loyalty may defect when national economic conditions are perceived as inadequate, especially among Democrats who believe their party is not managing the economy well. Table 2 outlines the coefficients of all six models. The primary findings remain robust, indicating the strong influence of economic hardships on voting behavior.

Table 2: Coefficients for all Models

| | Logit Models (Odds Ratio) | | | | | |
|---|---|---|---|---|---|---|
| | Primary DV: Vote for Incumbent ( 1 = Biden; 0 = Trump) | | | | | |
| **Variables** | **Model 1 (Pocketbook)** | **Model 2 (Pocketbook)** | **Model 3 (Pocketbook)** | **Model 4 (Sociotropic)** | **Model 5 (Sociotropic)** | **Model 6 (Sociotropic)** |
| **Inflation B. Index (Main IV)** | 0.823*** (0.043) | 0.866* (0.059) | 0.933 (0.067) | | | |
| **Dem No-Econ (Main IV)** | | | | 0.030*** (0.006) | 0.044*** (0.010) | 0.034*** (0.009) |
| **Gender** | | 1.074 (0.187) | 1.082 (0.188) | | 0.938 (0.195) | 0.926 (0.195) |
| **Logged Age** | | 0.975 (0.269) | 0.993 (0.275) | | 0.962 (0.299) | 1.033 (0.320) |
| **Educ** | | 1.285*** (0.088) | 1.296*** (0.090) | | 1.315** (0.123) | 1.318** (0.125) |
| **Family Income** | | 0.983 (0.031) | 0.983 (0.031) | | 1.036 (0.040) | 1.037 (0.040) |
| **Employment Status: Ref: All Other** | | | | | | |
| - Full-time | | 0.901 (0.229) | 0.900 (0.229) | | 0.961 (0.278) | 0.986 (0.284) |
| - Part-time | | 1.379 (0.449) | 1.379 (0.444) | | 1.332 (0.502) | 1.365 (0.519) |
| - Unemployed | | 1.395 (0.511) | 1.357 (0.488) | | 1.977 (1.011) | 2.020 (1.057) |
| - Retired | | 1.033 (0.293) | 1.022 (0.291) | | 1.860 (0.627) | 1.957 (0.682) |
| **Big City** | | 1.212 (0.272) | 1.200 (0.272) | | 1.127 (0.326) | 1.113 (0.321) |
| **Rural Area** | | 0.757 (0.186) | 0.766 (0.186) | | 1.063 (0.336) | 1.112 (0.360) |
| **Own Home** | | 0.712 (0.387) | 0.702 (0.383) | | 1.251 (0.590) | 1.254 (0.579) |
| **Rent Home** | | 1.266 (0.672) | 1.251 (0.667) | | 1.719 (0.802) | 1.739 (0.797) |
| **Democrat** | | 23.364*** (5.328) | 54.534*** (23.656) | | 12.949*** (3.725) | 9.177*** (2.796) |
| **Dem*IBD** | | | 0.636*** (0.111) | | | |
| **Dem*NoEcon** | | | | | | 3.131* (1.739) |
| **Observations** | 1176 | 1176 | 1176 | 1176 | 1176 | 1176 |
| **Log Likelihood** | -800.266 | -558.618 | -554.799 | -508.809 | -405.556 | -402.860 |
| **BIC** | 1614.672 | 1223.284 | 1222.715 | 1031.759 | 917.160 | 918.838 |
| **AIC** | 1604.532 | 1147.236 | 1141.597 | 1021.619 | 841.112 | 837.720 |
| **Pseudo R-squared** | 0.010 | 0.309 | 0.314 | 0.371 | 0.498 | 0.502 |

- *p<0.05, ** p<0.01, *** p<0.001
- B (Se) = Coefficient (Standard Error)

I also run the "linktest" command after the regression models to ensure my regression models are well specified. The linktest output shows that all models are well-specified, as the _hatsq term is statistically insignificant. So, all the regression models are well-specified, and any model has no huge specification errors. The AIC and BIC thus give a basis for comparison of model fit between the Pocketbook and Sociotropic sets. For the Pocketbook models, Model 1 (Pocketbook Bivariate) has an AIC of 1604.532 and BIC of 1614.672; in this sense, it is the least fit among the Pocketbook models. When I control the sociodemographic characteristics of Model 2 for the pocketbook voting (Pocketbook Multivariate), both AIC and BIC values improve and become AIC = 1147.236 and BIC = 1223.284. Model 3 (Pocketbook with Interaction) further improved the model fit, as it has the lowest AIC of 1141.597 and a BIC of 1222.715 among the Pocketbook models, suggesting it has the best fit. For Sociotropic models, Model 4 (sociotropic bivariate) had AIC=1021.619 and BIC=1031.759, while Model 5 (sociotropic multivariate) displayed good improvement in these values, i.e., AIC=841.112 and BIC=917. Among sociotropic models, model 6 with Interaction has the lowest level of AIC, standing at 837.720. For its part, BIC reaches 918.838. Analyzing both AIC and BIC for six models, model 3 is the best fit for the pocketbook, and model 6 is the best fit for the sociotropic models.

In Model 1 (Pocketbook Bivariate), the coefficient on Inflation B. Index is 0.823* (0.043), meaning one's economic struggles increase, reducing the probability of voting for the Incumbent, which supports my hypothesis ($H_1$). Model 2 (Pocketbook Multivariate) also shows that controlling for the sociodemographic characteristics, as the degree of personal economic struggles increases, the likelihood to express intention to vote for the Incumbent decreases, and it is statistically significant (OR= 0.866*; P <0.5). In model 2, we also observe that both Education (OR = 1.285***; P < 0.001 ) and being self-identified as democrats (OR= 23.364*** ; P < 0.001) increase the likelihood of voting for the Incumbent, and both these factors are statistically significant. In Model 3 (Pocketbook with Interaction), the IBR becomes statistically insignificant, but now again, the Democrat coefficient is quite robust here at ( OR= 54.534*; P <0.05), which showcases the intention to vote along party lines. However, the most interesting finding in Model 3 is that the Interaction between being a Democrat and experiencing inflation-related behavior in the recent past significantly impacted the intention to vote along party lines. The coefficient for the interaction term DemIBD is (OR= 0.636**; P<0.01) and statistically significant. So, as personal economic hardship rises due to inflation in the past, the effect on voting for the Incumbent will dissipate for Democrats, so this result supports the hypothesis ($H_{1a}$). The level of Education remains significant ( OR= 1.3 ***; P<0.001), meaning that more Education will still increase the tendency to express their intention to vote for the Incumbent.

We still observe a similar tendency when we move to the individuals' more macro-level perception (sociotropic economic voting). Model 4 shows that (Model 4 Sociotropic Bivariate) the coefficient (OR = 0.03***; P<0.001) for individuals who perceive that the Democratic Party is not handling the economy and cost of living well are overwhelmingly less likely to express intention to vote for the incumbent president for the 2024 November re-election and it is statistically significant. Therefore, the bivariate logit model 4 capturing sociotropic economic voting showcases that the negative perception of the national economy decreases voting intentions for the Incumbent and supports my hypothesis ($H_2$).

In Model 5 (Sociotropic Multivariate), we observe similar trends for individuals who perceive that the Democratic Party is not handling the economy and cost of living well and are overwhelmingly less likely to express intention to vote for the incumbent president for the 2024 November re-election, and it is statistically significant (OR = 0.04***; P<0.001). So, this supports my hypothesis ($H_2$). Also, we observe in Model 5 similar to Model 2 (Pocketbook Voting-multivariate-with control variables), that we also observe that both Education (OR = 1.285**; P < 0.01 ) and being self-identified as democrats (OR= 23.364*** ; P < 0.001) increase the likelihood of voting for the Incumbent, and both these factors are statistically significant, meaning more schooling and party affiliation will be more likely to support the Incumbent for re-election.

Model 6 (Sociotropic multivariate with Interaction) also shows that individuals who perceive that the Democratic Party is not handling the economy and cost of living well are overwhelmingly less likely to express intention to vote for the incumbent president for the 2024 November re-election and it is statistically significant (OR = 0.034***; P<0.001). Similar to earlier multivariate models, the huge Democrat coefficient of 9.177* (2.796) reveals strong party loyalty. The interaction term, Dem*NoEcon = 3.131* (1.739), indicates positive signification, meaning Democrats with a negative sociotropic perception of the national economy are more probable to vote for the Incumbent; hence, strong party loyal voters do not accept hypothesis ($H_{2a}$). There is still significance within the Education level at the 0.05 level, with an odds ratio of 1.318 (0.125); in other words, higher Education increases the chances of voting for the Incumbent.

I control Gender, Logged Age, Family Income, Employment Status, Big City, Rural Area, Own Home, and Rent Home across the models. Yet none of these controls is statistically significant in any of the models. Models 1 and 2 support hypotheses ($H_1$) and ($H_{1a}$). Here, increasing personal economic hardships decrease the likelihood of voting for an incumbent. H1a (Pocketbook interaction) receives partial evidence through Model 3, as seen by the fact that personal economic hardships have significantly reduced effects on Democrats. H2 (Sociotropic effect) can be seen in Model 4, Model 5, and Model 6, wherein worse evaluations of the national economy decrease the likelihood of voting for an incumbent. In line with not supporting sociotropic Interaction, Model 6 shows that Democrats are more likely to vote for the Incumbent even when they hold negative perceptions of the national economy, indicating a strong party line. The significance of the level of Education can be noted in Models 2, 3, 5, and 6 since an increase in Education increases the likelihood of voting for the Incumbent.

### Pocketbook Voting and Inflation Behavior Index

Earlier research studies have significantly explored the effect of pocketbook voting in general elections and operationalized various techniques to ensure a set of both measures and methods to study the concept that voters are making choices based systematically on their pocketbook economic behavior. Studies have used self-interested economic perceptions (Cowden and Hartley 1992), economic perceptions  (Sigelman, Sigelman, and Bullock 1991), household income (Grafstein 2009), and changes in personal econometric status (Tilley, Garry, and Bold 2008). In my current study, I propose a composite Inflation Behavior Index to operationalize cumulative personal financial struggles and capture the effect of pocketbook voting behavior. This composite index gauges economic difficulties experienced over the past year in four binary variables related to inflation: whether one has borrowed money to pay bills, cut back on everyday spending, canceled or postponed major purchases, and dipped into savings. The index ranges between 0 and 4 regarding the level of inflation-affected behavior. The index was validated by Cronbach's Alpha 0.5672 (albeit very moderate, yet strong sequential factor loading) and confirmed by EFA and CFA, single-factor structure, and good factor loadings with a variance explanation of 43.64%. This unexplored method gives a full and robust measure of the impact of economic hardship on voting behavior; therefore, it extends the existing research by providing a detailed operationalization of pocketbook-voting tendencies. I ran five different bivariate logistic models with each behavior and index to explore the effect each inflation-induced behavior had last year and the inflation behavior index (IBR).

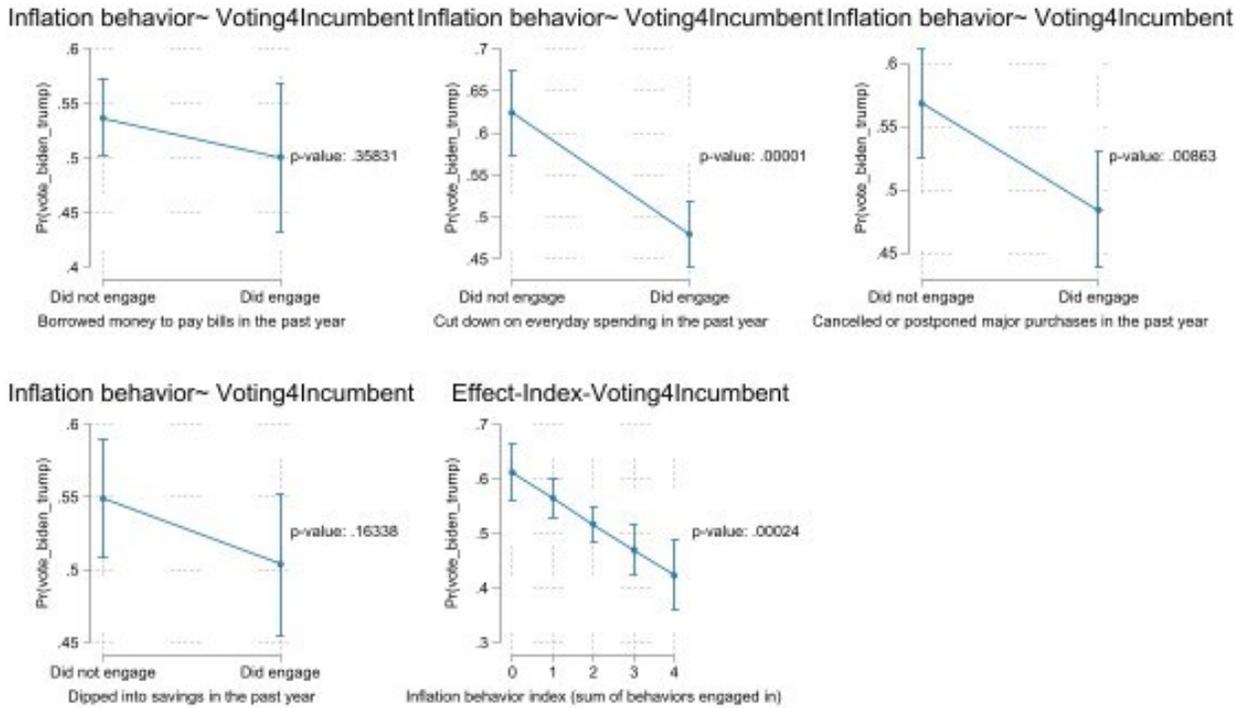

Figure 1: Single behavior vs Composite Behavior (Predictive Margins)

Here in Figure 1, I explore whether the binary composite index of the four binary variables (borrow_money, cut_spending, cancel_purchases, and dip_savings) is helpful and rational to operationalize for measuring the cumulative economic behavior of the respondents and their intended voting behavior in the forthcoming general election. I run four individual logit regression models with an odds ratio and then run the predictive margins from graphs 1 to 4, capturing the effect of borrow_money and dip_savings, which are not statistically significant with p-values of 0.358 and 0.163, respectively. However, cut_spending and cancel_purchases show statistically significant values at $p < 0.001$ and 0.01. The composite index (inflation_behavior_index) is statistically significant at a 0.002 probability level in predicting the dependent variable vote_biden_trump. This significance suggests that the combined effect of the four behaviors is highly predictive of voting intention. The odds ratio in the composite index is 0.82, with a confidence interval not containing 1, verifying its significance.

The rationale for using a composite economic behavior index to capture the effects of economic pocketbook voting in general elections is to increase the model's predictive power by seizing a broader construct than if just one had been used to make predictions, in this case, economic behavior during inflation. That is apparent with two individual behaviors that are not shown to be statistically significant but reflected as a significant composite index so that the combined impact should mean something. The pseudo R2 for the composite index model is 0.0097, improving the model fit of most variables added separately into the equation. Thus, the composite index provides a statistically significant predictor for understanding voting intentions and helps understand broader economic behaviors affecting voting decisions.

A composite index makes a model more parsimonious and simpler in interpretation because it essentially reduces the number of predictors while keeping important fundamental information. Therefore, the composite index of these four binary variables in the Logit model is commonly used and functional, even though the other two individual dummy variables are not statistically significant. With the composite index's significance, its categorization captures variability in the dependent variable vote_biden_trump in a meaningful manner; it thus seems to be a rational and logical construct that boosts explanatory power in this model.

### **Economic Behavior Index, Education, and Voting Intention**

Model 3 captures the interaction effects of being a self-identified Democrat and experiencing various degrees of inflation-related economic behavior and how that might impact the likelihood of expressing the intention to vote in the upcoming general election. In model 3, the main DV (economic behavior index) is not statistically significant anymore, but the coefficient for Education, self-identified Democrat, and the interaction variable are statistically significant. Figure 2 showcases the predictive margins of likelihood to express intention to vote for the Incumbent in the upcoming election based on the level of attained Education, being identified as a Democrat, and the interaction variable.

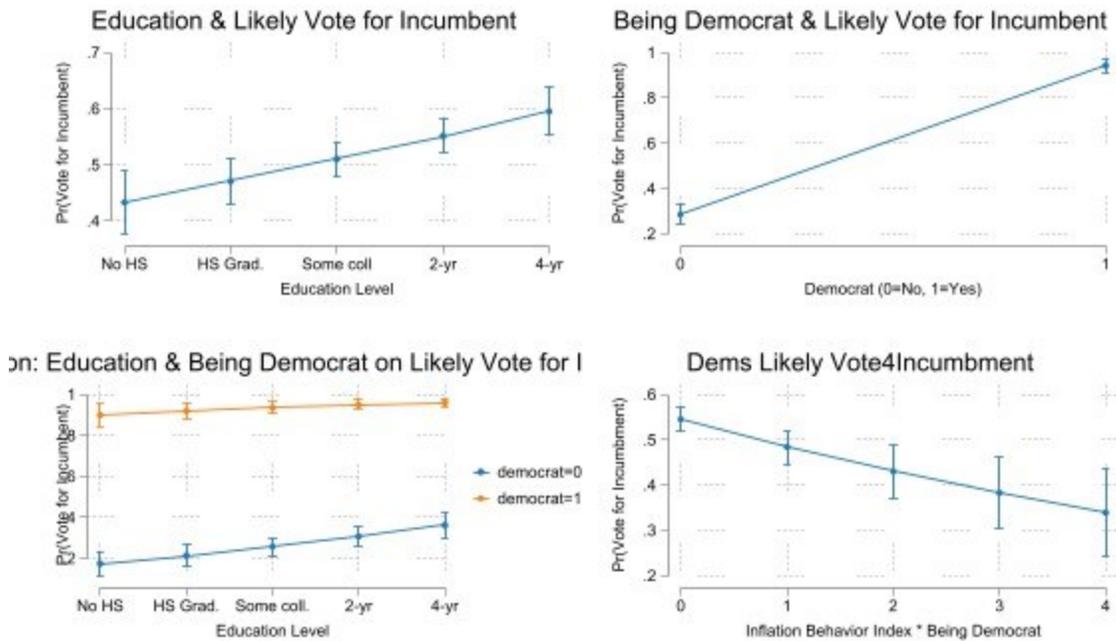

Figure 2: Pocketbook Economic Behavior and Likelihood to Vote for the Incumbent

The four predictive margins subplot in Figure 2 clearly show how Education and being a Democrat affect voting for the Incumbent, along with the Interaction between Education and experiencing inflation behavior on voting behavior. These insights directly support my hypotheses on pocketbook voting and party loyalty. In the first subplot, "Education & Likely Vote for Incumbent," higher education levels are linked to a greater likelihood of voting for the Incumbent. The predicted probability of voting for the Incumbent is low for those without a high school education. This probability steadily rises with higher Education, peaking for individuals with a 4-year degree. This suggests that more educated voters may be more resilient to personal economic hardships, more supportive of the Incumbent's policies, and more likely to express a clear intention to vote for the Incumbent. Similarly, the second subplot, "Being Democrat & Likely Vote for Incumbent," illustrates how being a Democrat increases the likelihood of voting for the Democratic party's Incumbent. Regardless of other variables, Democrats are far more likely to vote for the Incumbent than non-Democrats. This affirms that party identification is important in voting behavior since Democrats exhibit substantial support for the Democratic party incumbent.

The third subplot, "Interaction: Education & Being Democrat on Likely Vote for Incumbent," demonstrates the cumulative impact of Education and being a Democrat. It indicates that Democrats, regardless of educational level, are highly inclined to vote for the Incumbent. Higher Education marginally boosts non-Democrats' likelihood of voting for the Incumbent, but their total probability remains lower than that of Democrats. This implies that, while Education impacts voting behavior, party allegiance among Democrats is the more important element. The fourth plotline, "Dems Likely Vote4Incumbent," addresses the relationship between inflation behavior and being a Democrat. It shows that even among Democrats, the chance of voting for the Incumbent decreases when personal economic troubles, as assessed by the inflation behavior index, rise. This supports the theory that personal economic hardships might diminish the chance of voting for the Incumbent, even among strong party supporters ($H_{1a}$).

The four margin subplots in Figure 2 support my hypothesis ($H_1$ and $H_{1a}$). Higher educational attainment and self-identification as a Democrat increase the chance of voting for the Incumbent, but personal financial difficulties may diminish this likelihood, even among party loyalists. Personal economic realities and experiences have a significant influence on voter behavior. They may outweigh party loyalties, leading some self-identified Democrats to vote against party lines when confronted with rising inflation-related issues.

### Sociotropic Economic Voting Behavior

The collection of the first three models outlines that potential voters are less likely to vote for the Incumbent in the upcoming election when they face various degrees of economic struggles due to inflation. To a certain extent, even self-identified Democrats who are likely to vote along the party line tend to express against the party line when faced with increasing economic struggles due to ongoing inflation in the recent past. Cumulative personal economic struggles tend to outweigh the intention to vote along party lines, and it is statistically significant.

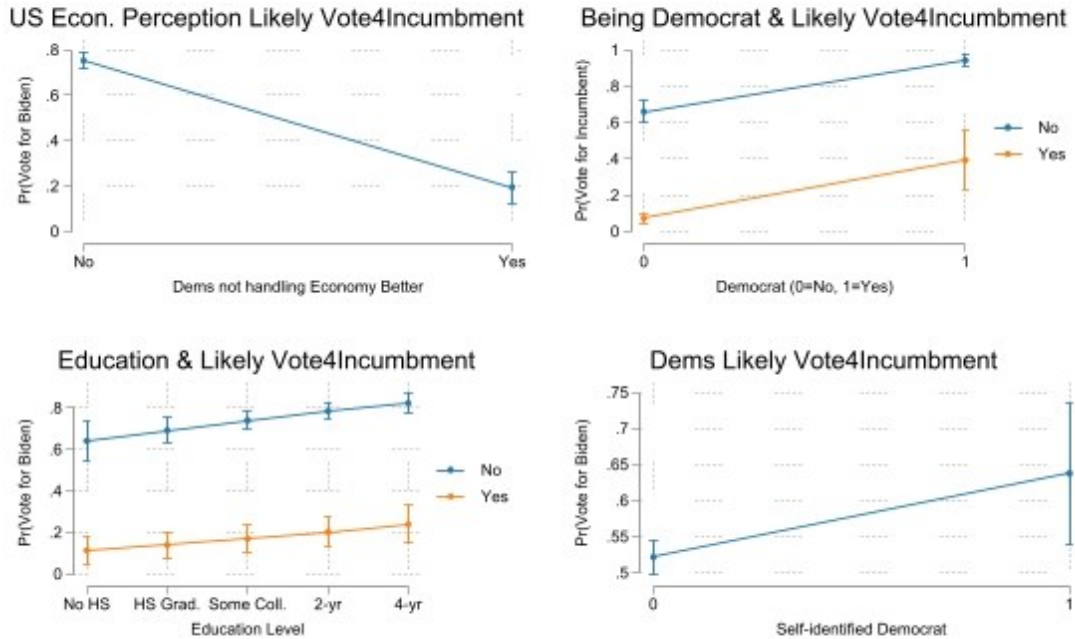

Figure 3: Sociotropic Economic Voting Behavior

Similar to Models 1, 2, and 3 (which capture pocketbook voting), Models 4, 5, and 6 capture the effect of sociotropic voting among potential U.S. voters in the upcoming 2024 general election. Both in Model 5 and Model 6, the main DV, Education, being a democrat, have statistically significant coefficients. The interaction variable between being a Democrat and thinking the Democratic party is not handling the national economy well is also included to capture the potential to vote against the party line (similar to model 3). Subplot one clearly shows that in a bivariate logistic model (odds ratio), when someone perceives that the incumbent party is less likely to handle the economy better and the rising cost of living, the likelihood of voting for the Incumbent significantly decreases. The second subplot signifies that a self-identified Democrat thinks the Democratic Party is handling the economy well (Blue Line; No) and is much more likely to vote for the Democratic Incumbent for re-election in November 2024 than those who believe otherwise (Orange Line: Yes). Similarly, those with higher levels of Education think the Democratic Party is handling the economy well (Blue Line; No) and are much more likely to vote for the Democratic Incumbent for re-election in November 2024 than those who believe otherwise (Orange Line: Yes).

Finally, the fourth and final subplot captures the Interaction between being a Democrat who still thinks the Democratic party is not handling the economy and cost of living well, being likely to vote along party lines, and expressing their intention to vote for the Incumbent in the upcoming general election. The key finding between model 3 (Pocketbook) and model 6 (sociotropic) is that personal economic struggles tend to defect party line voting. In contrast, the sociotropic tends to hold the party line voting, albeit on a much smaller scale.

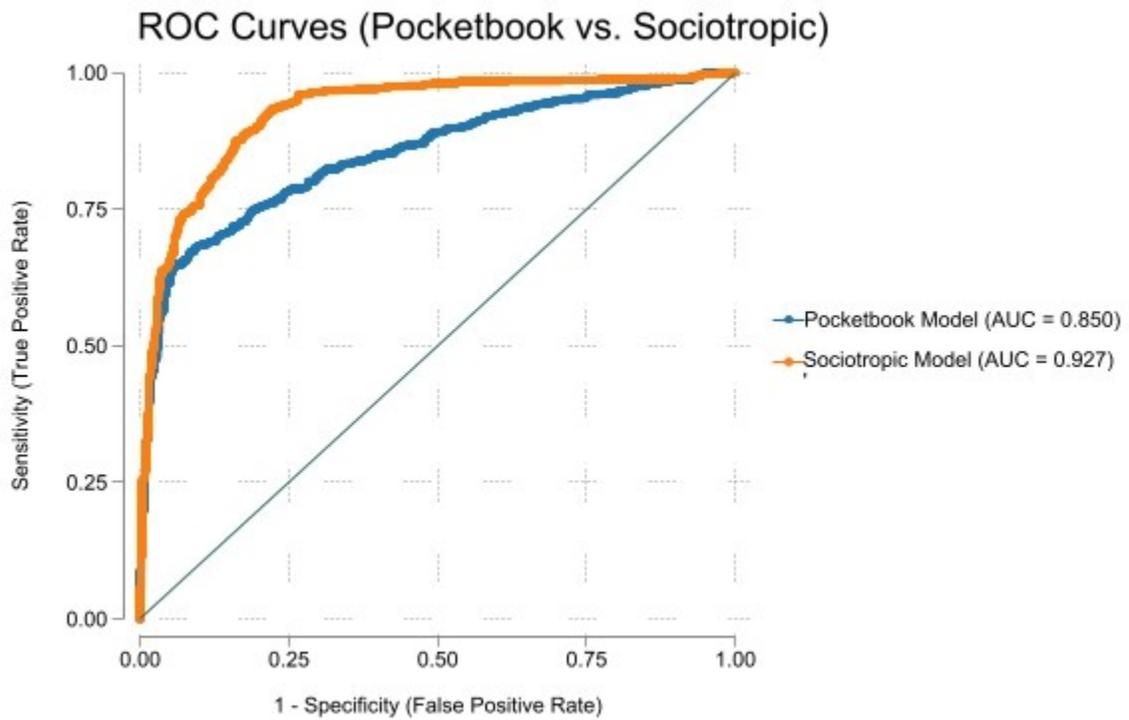

Figure 4: ROC Curves Predicting Likelihood of Voting for Incumbent, U.S. Presidential Election 2024

Figure 3 illustrates the ROC curves for the Pocketbook and Sociotropic models, highlighting their predictive power for the likelihood of voting for the Incumbent in the 2024 U.S. Presidential Election. The Area Under the Curve (AUC) values are .85 for the Pocketbook model and .92 for the Sociotropic model. The ROC curve demonstrates that personal financial struggles (pocketbook voting; full model 3) and perceptions of the broader economy (sociotropic voting; full model 6) influence voter behavior. The higher AUC for the Sociotropic model suggests that perceptions of the national economy might substantially impact voting decisions.

However, both models underscore the importance of economic factors in determining voter loyalty and behavior.

## Calibration Plots for Pocketbook and Sociotropic Models

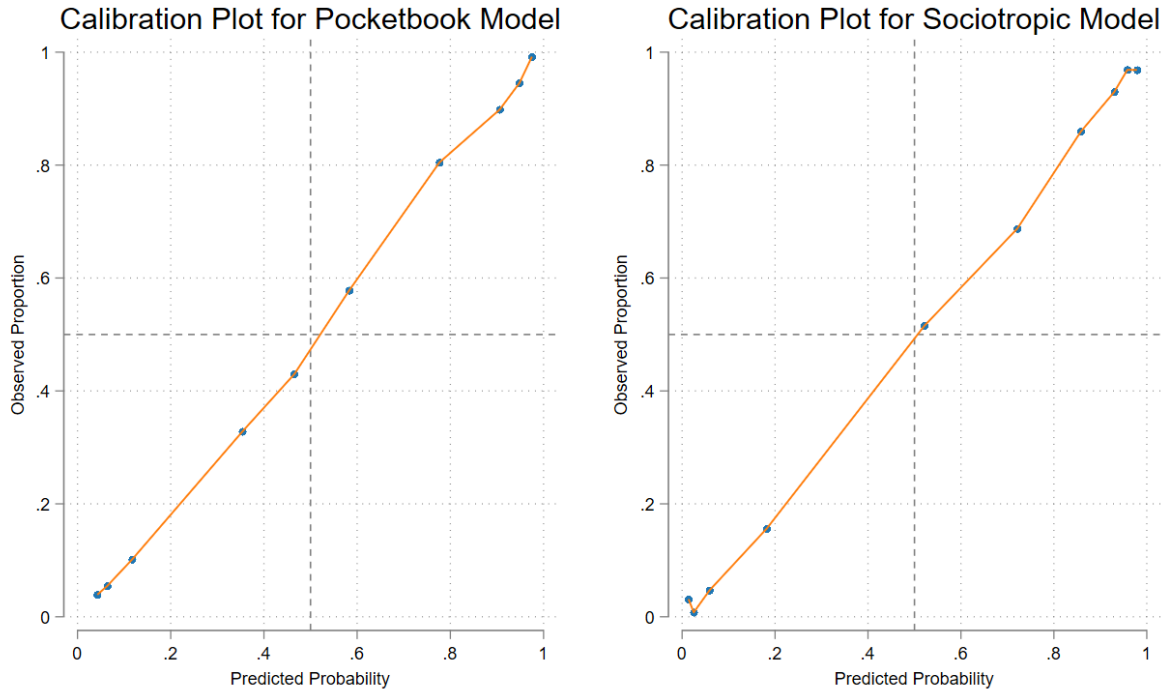

Figure 5: Calibration Plots for Pocketbook and Sociotropic Economic Voting Models

Figure 5 presents the calibration plots for both Pocketbook and Sociotropic models. These plots show how well the predicted probabilities from the models align with the observed proportions of voters likely to support the Incumbent. The calibration plot for the Pocketbook model shows that the predictions closely match actual voting intentions across multiple probability bins, indicating effective model calibration. Similarly, the sociotropic model's calibration plot shows accurate predictions consistent with observed data. These plots demonstrate the models' capacity to estimate voter behavior based on economic views and personal financial experiences, verifying the methodology and stressing the significance of economic considerations in voting decisions. In the current research study, I leverage the ANES 2024 Pilot Study data to examine the dynamics of economic voting within the context of the upcoming 2024 U.S. presidential election.

I only use the pilot study cross-sectional dataset from the ANES survey, so the generalizability and causality of the findings are very limited.  However, the current study emphasizes three things. I introduce a Cumulative Inflation-related behavior Index (IBR) that profoundly affects voters and even has the likelihood of breaking party loyalty in general elections. Secondly, voters tend to react less and vote against the party line when the economic issue is sociotropic. Still, when they cumulatively experience inflation-related financial struggles, they will likely vote against the party line. Finally, the level of attained Education has a nuanced effect on intention to vote in general elections; voters tend to vote along party lines, and non-democrats with at least four years of degree have a slight increase in intention to vote for the Incumbent.

In a highly polarized sociopolitical environment, my current study (a work in progress, and I am looking for feedback and collaboration to refine my work) contributes to the gap in existing knowledge by introducing an Inflation Behavior Index (IBR). I also acknowledge that the very moderate Cronbach's Alpha of 0.5672, compensated by validation through EFA and CFA, shows strong factor loadings, affirming its construct validity. Campaign strategists and political parties could focus significantly on ensuring the economic well-being of their constituents through various measures so that the party line voting doesn't break. Secondly, economic well-being and access to better Education positively affect voting for incumbent re-election. Finally, a more focused effort from the ANES data collection committee to capture more nuanced variables to capture and analyze the effects of economic voting would be highly influential and deeply meaningful in advancing voting behavior and public opinion research in the future.

<u>**Economic Struggles and Inflation: How Does that affect voting decision?**</u>

Muhammad **Hassan** Bin Afzal
Department of Political Science and Public Service
University of Tennessee at Chattanooga
WPF218@Tennessee.edu

**https://orcid.org/0000-0001-8192-0885**

**(Supplementary Materials: Codebook)**

## Introduction

I used data from the ANES 2024 Pilot Study to explore the dynamics of economic voting in the upcoming 2024 U.S. presidential election. This dataset, collected by the American National Election Studies (ANES), includes responses from approximately 1,779 participants and provides comprehensive insights into various economic and political behaviors. The survey was conducted online between February 20 and March 1, 2024, employing a randomized sampling process to ensure representativeness. Weighted responses were used to generalize findings to the U.S. adult population, making the study's results robust and reliable (ANES 2024). The subsequent sections include detailed summary statistics and descriptions of each variable used in my analysis.

**Summary Statistics Table:**

| Variable | Categories | Frequency | Percent | Description |
|---|---|---|---|---|
| **Inflation Behavior Index** | 0 behaviors (0) | 239 | 18.41 | No inflation response. Represents the sum of inflation-related behaviors engaged in. (IV for Model 1) |
| | 1 behavior (1) | 356 | 27.43 | One inflation-related behavior engaged in. (IV for Model 1) |
| | 2 behaviors (2) | 311 | 23.96 | Two inflation-related behaviors engaged in. (IV for Model 1) |
| | 3 behaviors (3) | 269 | 20.72 | Three inflation-related behaviors engaged in. (IV for Model 1) |
| | 4 behaviors (4) | 123 | 9.48 | All four inflation-related behaviors engaged in. (IV for Model 1) |
| **Gender** | Male (0) | 630 | 48.54 | Binary variable indicating respondent's gender. (Control) |
| | Female (1) | 668 | 51.46 | |
| **Education** | No H.S. credential (0) | 45 | 3.47 | A categorical variable representing the highest level of Education. (Control) |
| | High school graduate (1) | 387 | 29.82 | |
| | Some college (2) | 255 | 19.65 | |
| | 2-year degree (3) | 144 | 11.09 | |
| | 4-year degree (4) | 308 | 23.73 | |
| | Post-grad (5) | 159 | 12.25 | |
| **Family Income (Recoded)** | Less than $10,000 | 85 | 6.55 | Categorical variable representing family income range. (Control) |
| | $10,000 - $19,999 | 98 | 7.55 | |
| | $20,000 - $29,999 | 168 | 12.94 | |
| | $30,000 - $39,999 | 110 | 8.47 | |
| | $40,000 - $49,999 | 110 | 8.47 | |
| | $50,000 - $59,999 | 106 | 8.17 | |
| | $60,000 - $69,999 | 94 | 7.24 | |
| | $70,000 - $79,999 | 116 | 8.94 | |
| | $80,000 - $99,999 | 106 | 8.17 | |
| | $100,000 - $119,999 | 96 | 7.40 | |
| | $120,000 - $149,999 | 85 | 6.55 | |
| | $150,000 and above | 124 | 9.55 | |
| **Race (Combined)** | Native American/Two or more races/Other (1) | 61 | 4.70 | Categorical variable representing respondent's race. (Control) |
| | White (2) | 893 | 68.80 | |
| | Black (3) | 155 | 11.94 | |
| | Hispanic (4) | 159 | 12.25 | |
| | Asian (5) | 30 | 2.31 | |
| **Employment Status** | Other (0) | 262 | 20.22 | Categorical variable representing employment status. (Control) |
| | Full-time (1) | 494 | 38.12 | |
| | Part-time (2) | 157 | 12.11 | |
| | Unemployed (3) | 93 | 7.18 | |
| | Retired (4) | 290 | 22.38 | |
| **Lives in a Big City** | No (0) | 1,030 | 79.35 | Binary variable indicating residence in a big city. (Control) |
| | Yes (1) | 268 | 20.65 | |
| **Lives in a Rural Area** | No (0) | 1,077 | 82.97 | Binary variable indicating residence in a rural area. (Control) |
| | Yes (1) | 221 | 17.03 | |
| **Owns Home** | No (0) | 402 | 30.97 | Binary variable indicating home ownership. (Control) |
| | Yes (1) | 896 | 69.03 | |
| **Rents Home** | No (0) | 945 | 72.80 | Binary variable indicating if the respondent rents their home. (Control) |
| | Yes (1) | 353 | 27.20 | |
| **Party Identification** | Other/Not sure (0) | 104 | 8.13 | Categorical variable representing party identification. (Control) |
| | Democrat (1) | 455 | 35.55 | |
| | Republican (2) | 383 | 29.92 | |
| | Independent (3) | 338 | 26.41 | |
| **Democratic Party Handling** | No (0) | 771 | 59.40 | The binary variable indicates if the respondent believes the Democratic party is not handling the national economy better. (IV for Model 2) |
| | Yes (1) | 527 | 40.60 | |

| Interaction Variables | Democrat * Inflation Behavior Index | See index | - | Interaction of Democrat identification with the Inflation Behavior Index. (Mediator for Model 1) |
|---|---|---|---|---|
| | Dems not handling econ * Democrat. | See above | - | Interaction of Democrat identification with the perception that Dems are not handling the economy. (Mediator for Model 2) |

**Inflation Behavior Index**

- **Construction**: The index is constructed from four variables from the ANES 2024 Pilot Study, measuring behaviors such as borrowing money to pay bills, cutting down on everyday spending, canceling or postponing major purchases, and dipping into savings. Each behavior is coded as 1 if engaged and 0 otherwise, resulting in an index ranging from 0 to 4.
- **Summary**:
  - 0 behaviors (no inflation response): 18.41%
  - 1 behavior: 27.43%
  - 2 behaviors: 23.96%
  - 3 behaviors: 20.72%
  - 4 behaviors (all four responses): 9.48%

**Footnotes**

1. **DV (Dependent Variable)**: Vote for Biden/Trump
2. **IV (Independent Variable)**: Inflation Behavior Index, Dems not handling economy better
3. **Control**: Gender, Education, Family Income, Race, Employment Status, Residence in a big city, Residence in a rural area, Owns Home, Rents Home, Party Identification

4. **Mediator**: Democrat * Inflation Behavior Index, Dems not handling econ * Democrat

This table provides a detailed breakdown of the demographic and economic variables used in the analysis, emphasizing their distribution and role in the models.

## Further Explanation of the Inflation Behavior Index

I utilize four variables from the ANES 2024 Pilot Study to construct an index measuring inflation behavior over the past year. This pilot study, conducted by the American National Election Studies, involved a comprehensive survey of approximately 1,779 participants to gather detailed data on various economic and political behaviors. The specific variables I focus on are whether individuals borrowed money to pay bills (infl_behav_1), cut down on everyday spending (infl_behav_2), canceled or postponed major purchases (infl_behav_3), and dipped into savings (infl_behav_4). Each variable captures a different aspect of how people respond to inflationary pressures.

First, I clean the data by removing any cases where the response is marked as "inapplicable" or "legitimate skip" (coded as -1). For instance, in infl_behav_1, 6.81% of the responses are coded as -1, which I remove to ensure that my analysis includes only relevant answers. The remaining responses are recorded so that a response indicating that the individual engaged in the behavior (coded initially as 1) remains 1, and a response indicating that they did not engage in the behavior (coded initially as 2) is recorded as 0. After recoding, I create a new variable that sums the values of these four behaviors, resulting in an index ranging from 0 to 4. A score of 0 indicates that the respondent did not engage in any of the behaviors, while a score of 4 indicates that they engaged in all four.

I conducted several analyses to ensure this index's internal consistency and validity. The summary statistics show that the mean of the Inflation Behavior Index is approximately 1.73, with a standard deviation of about 1.25. The frequency distribution reveals that 19.22% of respondents did not engage in any of the behaviors, 27.82% engaged in one, 23.38% engaged in two, 20.07% engaged in three, and 9.50% engaged in all four. I then checked the internal consistency of the index using Cronbach's Alpha, which is 0.5672. Although this indicates moderate internal consistency, suggesting that the items are related and collectively provide helpful information about how individuals cope with inflation, I further validate the index through Exploratory Factor Analysis (EFA) and Confirmatory Factor Analysis (CFA).

EFA reveals that a single factor explains 43.64% of the variance with eigenvalues above 1, confirming that these behaviors cluster well. The factor loadings are 0.5596 for borrowing money, 0.7057 for cutting down on spending, 0.7447 for canceling purchases, and 0.6163 for dipping into savings, indicating solid relationships with the underlying inflation behavior factor. CFA further confirms this single-factor structure, with significant coefficients for all behaviors, indicating robust relationships and good model fit.

Despite the Cronbach's Alpha of 0.5672 being slightly below the ideal threshold, the high factor loadings in EFA and the confirmed structure in CFA demonstrate that the index is a valid tool for estimating public attitudes toward inflation-related behaviors. These analyses collectively validate the index, making it a reliable and valuable measurement tool for understanding how individuals respond to inflation.